# Investigation of pionic contribution in the lepton and anti-lepton production cross section in p-Cu and p-Pt collision

R. Rostami; F.Zolfagharpour;

*Department of Physics, University of Mohaghegh Ardabili, Ardabil, Iran*

Email: rezarostami62@gmail.com

Email: zolfagharpour@uma.ac.ir

**Abstract**

*For detailed explanation of the experimental results of lepton production cross section in hadronic collisions such as nucleon-nucleon or nucleon-nuclei, it is of great importance to use quarks and sea quarks distribution function inside free and bound nucleons. In this paper the role of pion cloud inside the nucleus in the structure function of Cu and Pt nuclei and the EMC ratio of these nuclei were investigated by using harmonic oscillator model. Then in the framework of the Drell-Yan process and conventional nuclear theory, GRV's quarks distribution functions and pionic quarks distribution functions were used to calculate lepton and anti-lepton production cross section in p-Cu and p-Pt scattering. From the results and based on the mentioned model, by considering pionic contribution, the theoretical results are improved.*

Introduction

Nuclei are made up of protons and neutrons bounded by strong nuclear forces, with the binding energy much less than the rest mass of the nucleons. Therefore, it is expected that scattering cross section of the leptons from nucleus be obtained from non-coherent summation of scattering cross sections from individual bound nucleons[1-5]. Hence this impulse approximation seems to explain the experimental data obtained from relativistic lepton-nucleus or hadronic collision. However, in 1983, the experiment done by a group of European researchers confronted scientists with experimental results which demonstrated that the structure of free and bound nucleons was different[6-8]. It was therefore concluded that some phenomena existed which were responsible for the slight difference in the internal structure of nucleons and consequently in quarks distribution functions in the bound nucleons compared to free nucleons. This change in quark distribution functions, which alters the free nucleons structure functions, is known as the EMC effect. The researches done by many other groups, such as Thomas and his colleagues in the following years[1,9,10] showed that pure fermi motion cannot explain the EMC effect and that binding energy plays an important role. Their effect is the result of nuclear interactions which play an important role in the deep inelastic scattering of leptons from nuclei. However it should be mentioned that the consideration of these phenomena alone is insufficient to explain the difference between quarks distribution functions inside bound and free nucleon. Ericson and Thomas[10] investigated the role of the pion field in the enhancement of the EMC effect in iron. Simona Malace et al [10] performed an overview of the EMC effect. In this paper, by using the works of mentioned authors and others [3,9-13]to improve on previous results, the effect of Fermi motion and binding energy and pionic contribution was considered in extracting the quarks distribution functions inside the bound nucleon in the Cu and Pt nucleus,while other phenomena like shadowing effect and quark exchange were disregarded[9].Therefore, the dileptons production cross sectionin $p + nucleus \to l^+ + l^- + x$ scattering can be calculated by using the mentioned effects in the framework of conventional nuclear theory and Drell-Yan process. The used model is the harmonic oscillator model that is modified by considering different $\hbar\omega$ parameters for occupied nucleon energy state inside nucleus. This modification is done since nucleons in deferent state may be affected by different main field. In the case of less binding energy, better agreement between the theoretical results and experimental data can be achieved compared with the case when the nucleons are affected by fixed main field [14-17].GRV's quarks distribution functions for bound nucleons and pions are used for the calculation of quarks distribution inside nucleus[11]. In this work, to investigate the pionic contribution in the lepton and anti-lepton production cross section in p-Cu and p-Pt collision, by first considering the pionic contribution, the structure function of these nuclei and their corresponding EMC ratio were studied. Then their pionic contribution in quarks distribution functions was probed and finally the amount of this contribution in the dilepton production cross section in p-Cu and p-Pt scattering was computed.

**Calculation of the structure function and EMC ratio of the $^{63}$Cu, $^{195}$Pt nuclei**

The nucleus structure function by considering the pion cloud effect is defined as follows[12,17]:

$$F_2^A(x, Q^2) = \int_x^A f_\pi^A(z) F_2^\pi\left(\frac{x}{z}, Q^2\right) dz + \sum_{N=n,p} \sum_{nl} \int_x^A dz g_{nl}^N f_N^A(z)_{nl} F_2^N\left(\frac{x}{z}\right) \quad (1)$$

Where the first term indicates the pionic contribution and the next term indicates the nucleon contribution in the nucleus structure function. $z = \frac{p_{nl}q}{m_N q_0}$ is the fraction of the total momentum carried by the nucleons and $x = \frac{Q^2}{2M_N q_0}$ is Bjorken scaling variable. $F_2^\pi\left(\frac{x}{z}, Q^2\right)$ is the pion structure function.

The distribution function of pion inside the nucleus is given as follows [17]:

$$f_\pi^A(z) = \frac{3g^2}{16\pi^2} \Delta \lambda z \left[\frac{1}{\lambda} \exp\left[-2\lambda \frac{t_0 + m_\pi^2}{m_\pi^2}\right] + \frac{1}{2} E_i\left[-2\lambda \frac{t_0 + m_\pi^2}{m_\pi^2}\right]\right] \quad (2)$$

Where $E_i(-z) = -\int_z^\infty dt \frac{e^{-t}}{t}$, $t_0 = \left|m_N^2 \frac{z^2}{1-z}\right|$ and g= 13.5 is the coupling constant.

The cut off parameter λ plays the most substantial role. When the nucleon is embedded in a nucleus, several modifications, such as the polarization of nuclear medium, occur. These may expressed be by an effective change of λ. The nuclear binding energy causes a violation of sum rule. This violation is proportional to the factor $<z>_N$ which is defined as follows[14]:

$$<z>_N = \frac{1}{A} \int_x^A dz\, z\, f^A(z) = 1 + \frac{<\varepsilon_\lambda>}{m_N} \quad (3)$$

Where $<\varepsilon_\lambda>$ in [12] is the mean one nucleon separation energy or on the other hand, is the average removal energy. But in [14-17] $<\varepsilon_\lambda>$ is considered to be different for nucleons in different levels[14-17]. The nucleons carry only a fraction $<z>_N < 1$ which in our calculation has been calculated as 0.9675 and 0.9707 for Cu and Pt nuclei, respectively. By considering pions and ignoring the contribution made by virtual particles like Δ particle and heavy mesons, the momentum sum rule can be written as follows:

$$<z>_N + \eta_\pi = 1 \quad (4)$$

Where $\eta_\pi$ is the momentum fraction carried by pions, which is defined as:

$$\eta_\pi = \int_0^{\frac{M_A}{m_N}} dz\, z\, f_\pi^A(z) \quad (5)$$

In the second term of equation (1), the first sum is related to the total number of neutrons and protons and the second is related to quantum number of each energy state. $g_{nl}^N$ is the occupation number of energy state $\varepsilon_{nl}$ such that for protons, N = p and for neutrons, N = n. $F_2^N(x)$ is the structure function of free nucleons in which $F_2^{N=n}(\frac{x}{z})$ and $F_2^{N=p}(\frac{x}{z})$ refer to neutron and proton, respectively.

Function $f_N^A(z)_{nl}$, which is calculated by the wave functions of harmonic oscillator, describes energy and momentum distribution of nucleons inside the nucleus as follows [12, 13]:

$$f_N^A(z)_{nl} = \frac{1}{2}\left(\frac{m_N}{\hbar\omega}\right)^{1/2} \frac{n!}{\Gamma(n+l+\frac{3}{2})} \sum_{t_1=0}^{n}\sum_{t_2=0}^{n} \frac{(-1)^{t_1+t_2}}{t_1+t_2} \binom{n+l+\frac{1}{2}}{n-t_1}$$
$$\times \binom{n+l+\frac{1}{2}}{n-t_2} \Gamma\left[l+t_1+t_2+1, \frac{m_N}{\hbar\omega}\left(z-1-\frac{\varepsilon_{nl}}{m_N}\right)^2\right] \quad (6)$$

Where $\varepsilon_{nl}$ is the energy of nucleon in state $n$, $l$ and $m_N$ is the mass of nucleon. $f_N^A(z)_{nl}$ should satisfy the normalization rule:

$$\sum_{N=n}\sum_{nl} \int_0^\infty dz\, g_{nl}^N f_N^A(z)_{nl} = A \quad (7)$$

Taking into consideration the sea quarks and gluons contributions, the structure function of nucleons satisfies the sum rules:

$$\int_0^1 F_2^N(x)dx = 1 \quad (8)$$

In the harmonic oscillator, $\hbar\omega$ in the natural unit can be expressed as:

٣

$$\hbar\omega(\text{MeV}) = \frac{42.2}{<r^2>_{nl}}\left(2n + l + \frac{3}{2}\right) \tag{9}$$

Where $<r^2>_{nl}$ is the mean square radius of state *n, l* and its unit is Fermi. The EMC ratio as the ratio of the structure function of nucleus to the deuterium structure function is defined as:

$$R_{\text{EMC}}(x) = \frac{2F_2^A(x,Q^2)}{AF_2^{2H}(x,Q^2)} \tag{10}$$

**Drell-Yan process**

In 1970, initial studies of lepton pair production $\mu^+\mu^-$ in Hadron-Hadron collision was reported by Christenson et al., [18] as follows:

$$h_A + h_B \rightarrow l^+l^- + x \tag{11}$$

The process of which is shown in Figure 1. The dilepton production cross section is decreased by increasing the dilepton mass. Drell and Yan [19] suggested that in Hadron-Hadron collision, leptons with opposite sign are produced. In the first stage, a quark of a hadron is annihilated with an antiquark of other hadrons and a virtual photon is generated. The generated photon is converted into a lepton pairs with opposite sign like $\mu^+\mu^-, e^+e^-$. This is an electromagnetic process which can be calculated. Many research groups such as Miller and his colleagues have investigated deep inelastic scattering for different purposes in the framework of the Drell-Yan process [20]. The role of different phenomena, including Fermi motion and binding energy, together with the pionic contribution in nuclear Drell-Yan process has been probed by various case studies[21-23]. In this paper, in order to improve on previous results, the pionic contribution in the dilepton production cross section in the framework of the Drell-Yan process and convention nuclear theory [3,9-13] was calculated by using harmonic oscillator model. In order to predict the differential cross section of lepton pairs production in an interaction like Figure 1, an idea of the quarks and anti-quarks distribution functions inside A and B hadrons was necessary. The lepton production cross section in the collision of two hadrons in terms of quarks and anti-quarks distribution functions is calculated as follows [24]:

$$\left(s\frac{d^2\sigma}{d\sqrt{\tau}dy}\right) = \frac{8\pi\alpha^2}{9\tau^{\frac{1}{2}}}\sum_i e_i^2 \left[q_i^A(x_1)\bar{q}_i^B(x_2) + \bar{q}_i^A(x_1)q_i^B(x_2)\right] \tag{12}$$

Where $\sqrt{\tau} = \frac{m}{\sqrt{s}} = \sqrt{x_1 x_2}$, $\sqrt{s}$ is the hadron-hadron c.m. energy and $m = \sqrt{x_1 x_2 s}$ is the mass of dileptons. $e_i$ is the fractional electric charge of quark *i*. $q_i^A(x_1)$ and $\bar{q}_i^B(x_2)$ are ith quark and anti-quark distribution function inside A and B hadrons, respectively. $q_i^A(x)$ is defined as [25]:

$$q_i^A(x) = \sum_{nl}\int_x^A \frac{dz}{z} g_{nl}^N f_N^A(z)_{nl} q_i\left(\frac{x}{z}\right) \tag{13}$$

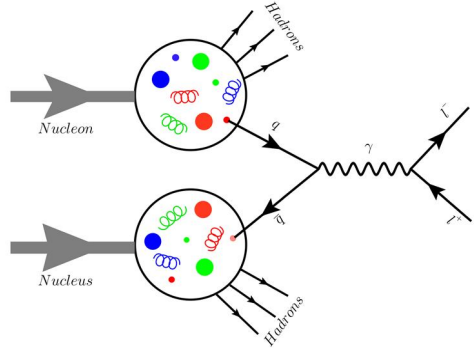

Figure 1: Drell-Yan process, a quark with fraction of momentum $x_1$ in the hadron A with an anti-quark with fraction of momentum $x_2$ in the hadron B collide and annihilate to create a photon. This virtual photon with mass $m = \sqrt{x_1 x_2 s}$ decays into a lepton pair [26].

**Results and discussion:**

In this work, the structure function of Cu and Pt nuclei were first calculated. For these calculations, the distribution function of nucleons inside nucleus was calculated by using the parameters listed in Table 1.

Table 1: Brackets contain ($< r^2 >^{1/2}$ (Fermi), $\hbar\omega$(MeV), $g_{nl}^p$, $g_{nl}^n$, $\varepsilon_{nl}$(MeV)) parameters for shells with quantum number n, l, respectively. $< r^2 >^{1/2}$ is taken from [27] for each level.

| shell | Nucleus | | |
|---|---|---|---|
| | $^2$H | $^{63}$Cu | $^{195}$Pt |
| 0s | (2.09,15.35,1,1,-1.5) | (1.67,22.23,2,2,-33) | (1.67,22.23,2,2,-34) |
| 0p | | (2.44,17.34,6,6,-32) | (2.44,17.34,6,6,-33) |
| 0d | | (3.10,12.51,10,10,-31) | (3.10,12.51,10,10,-32) |
| 1s | | (3.48,11.95,2,2,-30) | (3.48,11.95,2,2,-31) |
| 0f | | (3.95,11.92,9,14,-29) | (3.95,11.92,14,14,-30) |
| 1p | | | (4.44,11.39 ,6,6,-29) |
| 0g | | | (4.49,11.28,18,18,-28) |
| 1d | | | (4.55,10.98,10,10,-27) |
| 2s | | | (6.67,10.64 ,2,2,-26) |
| 0h | | | (5.15,10.32,8,22,-25) |
| 1f | | | (5.23,10.02,0,14,-24) |
| 2p | | | (5.41,9.37,0,6,-23) |
| 0i | | | (5.43,9.30,0,5,-22) |

Figure 2 shows the GRV's free neutron and proton structure function which were applied in the calculations. From equation (3), $< z >_N$ was calculated as 0.9675 and 0.9707 for Cu and Pt nuclei, respectively. Also From equation (4), $\eta_\pi$ was calculated as 0.0325 and 0.0299 for Cu and Pt nuclei, respectively. The values $\lambda = 0.026$, $m_\pi = 139.57$ were chosen and $\Delta\lambda$ had to be calculated for each nucleus. In Table 2, convenient $\Delta\lambda$ for each nucleus has been considered to be satisfactory for equation (5). Figure 3 shows a plot of the calculated distribution function of pion inside $^2$H, $^{63}$Cu and $^{195}$Pt.

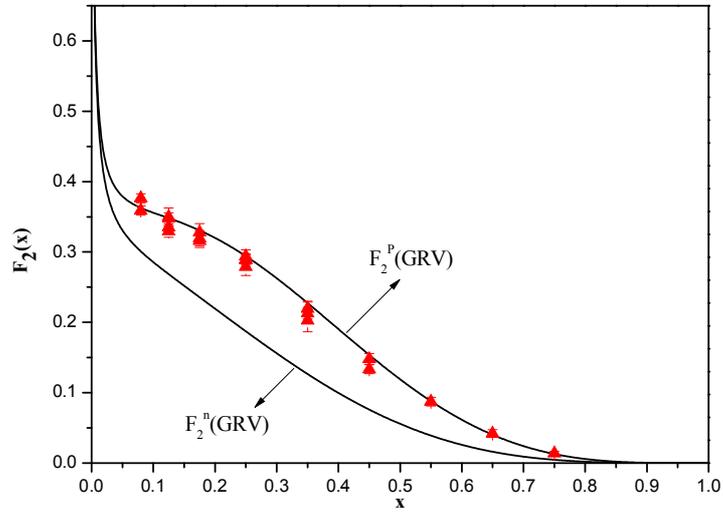

Figure 2: GRV's free neutron and proton structure function. Experimental data are taken from [28].

Table 2: The considered convenient Δλ for $^2$H, $^{63}$Cu and $^{195}$Pt nuclei in equation 2.

| nucleus | $^2H$ | $^{63}Cu$ | $^{195}Pt$ |
|---------|-------|-----------|------------|
| Δλ      | 0.00242 | 0.00792 | 0.00861 |

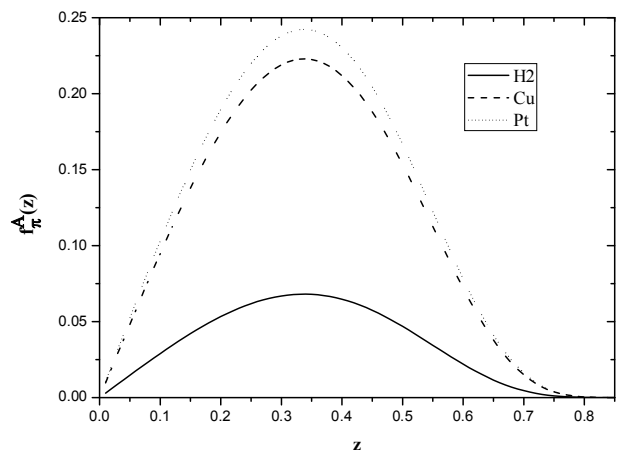

Figure 3: The distribution function of pion inside $^2$H, $^{63}$Cu and $^{195}$Pt nuclei.

In this work, the structure function of pion was been calculated from GRV's quark distribution functions. Figure 4 shows a plot of the structure function and the structure function calculated according to equation of $F_2^\pi(x)$ in [29].

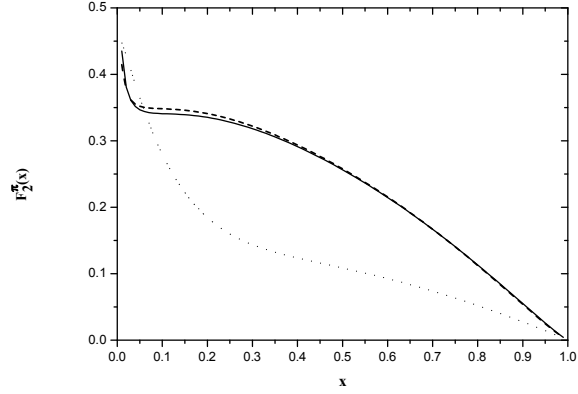

Figure 4: The structure function of pion. The full and dash curve are the GRV's LO and NLO pion structure function. The dot line is the structure function calculated according to equation of $F_2^\pi(x)$ in [29].

Figures 5 and 6 show plots of the structure function of Cu and Pt nuclei and the EMC ratio of these nuclei, by and without considering the pionic contribution. The dash line shows the structure functions of Cu and Pt nuclei and their corresponding EMC ratio by considering the Fermi motion and the binding energy according to $<\varepsilon_\lambda>$ which is listed in Table 1. According to references [12] and [14-17], the EMC ratio of some nuclei was plotted by considering $<\varepsilon_\lambda> = -40 MeV$ and different $<\varepsilon_\lambda>$ for nucleons of each level, respectively. The extracted results in references [14-17] are compatible with the experimental data.

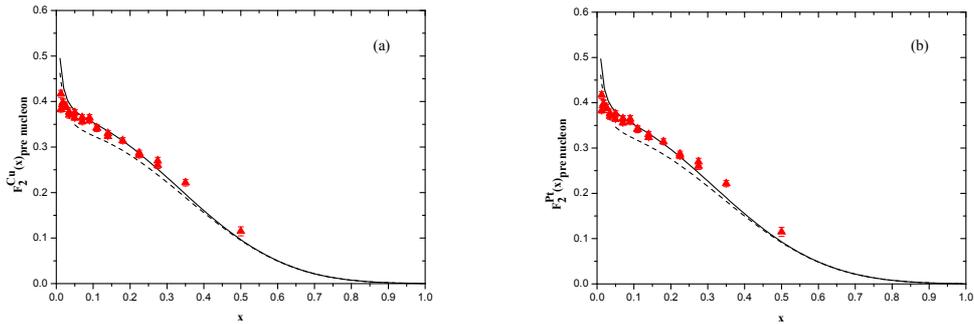

Figure 5: a and b show $^{63}$Cu and $^{195}$Pt structure functions per nucleon in mean $Q^2 = 5 GeV^2$. The full curve is obtained by considering, the Fermi motion, the binding energy, and the pionic contribution effect. The dash line shows the structure function by considering the Fermi motion, and the binding energy. Experimental data shows deuterium structure function per nucleon [28].

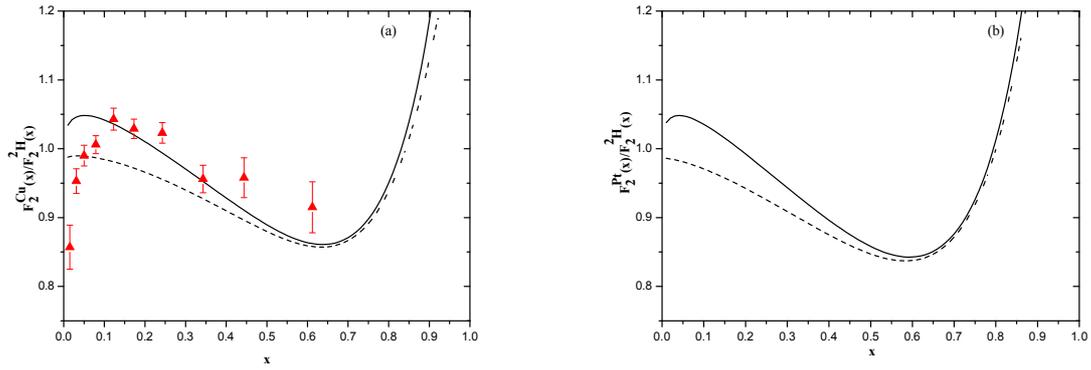

Figure 6: a and b show the ratio $R = \frac{F_2^A}{F_2^{2H}}$ in terms of x for Cu and Pt nuclei. The full curve shows the results by considering, the Fermi motion, the binding energy, and the pionic contribution effect. The dash line shows by considering the Fermi motion, and the binding energy. Experimental data are taken from [28].

The results showed that for x=0.15, pion cloud increased the structure function to about 7.9% for $^{63}$Cu and about 8.8% for $^{195}$Pt, respectively. Also, for x=0.15 and by considering pionic contribution, the EMC ratio of $^{63}$Cu and $^{195}$Pt nuclei were increased to 5.44% and 6.28%, respectively. According to reference [30], this contribution for deuteron and three-body nuclei was calculated to be less than 5%. To calculate pionic contribution in dilepton production cross section in hadronic collisions, it was necessary to first calculate quarks and sea quarks distribution functions inside nucleus by considering the pionic contribution. In Figures 7 and 8, by using GRV's quarks distribution functions in nucleons and pions, the quarks and sea quarks distribution functions inside $^{63}$Cu and $^{195}$Pt nuclei have been calculated.

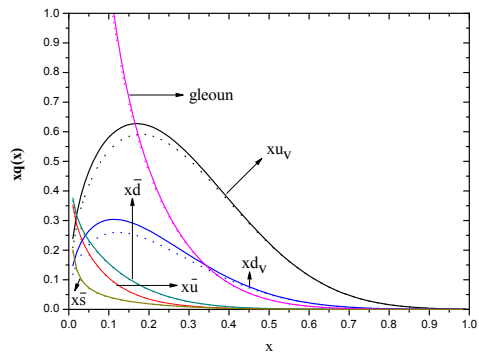

Figure 7: Distribution function of valance and sea quarks inside $^{63}$Cu nucleus at $Q^2 = 20 GeV^2$. The full curve is obtained by considering the pionic contribution and, the dash line is plotted without considering the pionic contribution.

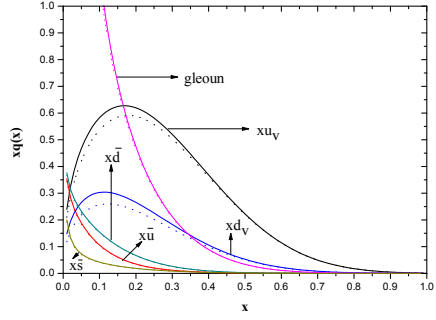

Figure 8: Distribution function of valance and sea quarks inside $^{195}$Pt nucleus at $Q^2 = 20 GeV^2$. The full curve is obtained by considering the pionic contribution and, the dash line is plotted without considering the pionic contribution.

Figure 9 shows dilepton production cross section in p-Cu scattering by and without considering the pionic contribution for y=-0.2,0,0.1, and 0.4, and by considering K factor. In Figure 10, by and without considering the pionic contribution, the lepton and anti-lepton production cross section in p-Pt collision for y=0.025, 0.163, and 0.6 have been plotted by considering K factor. The amount of used K factor is indicated in Figures 9 and 10. Results show that K factor increased linearly with increasing rapidity y. The linear fit of the K factor in terms of y for $^{63}$Cu and $^{195}$Pt nuclei is obtained as follows:

$$K = \frac{3}{4}y + 1.7 \qquad for\ Cu$$

$$K = 0.3y + 1.87 \qquad for\ Pt \qquad (14)$$

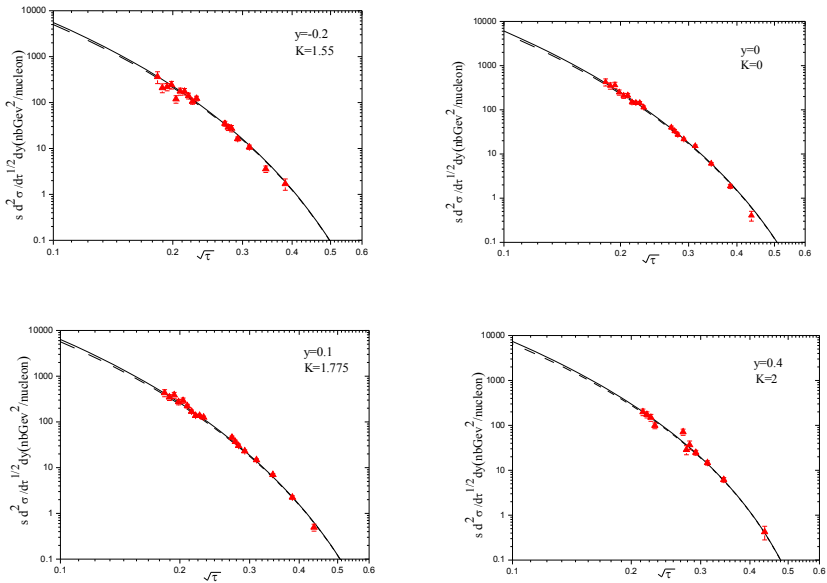

Figure 9: Dileptons production cross-section for p-Cu collision in terms of $\sqrt{\tau}$ according to Eq.13 for y = -0.2,0,0.1, 0.4 and $\sqrt{s} = 38.8$ GeV. The full curve is obtained by considering the pionic contribution and the dash line is plotted without considering the pionic contribution. The used K factor has been indicated in each graph. The difference between full and dash lines is about 3 - 8%. The experimental results have been taken from [28].

According to reference [24] different amount for K factor has been reported by many research group.

Pionic cloud resulted in an increase in the dileptons production cross-section to about 3 -8 percent in p- Cu collision and about 3- 6 percent in p- Pt collision in available experimental data ranges.

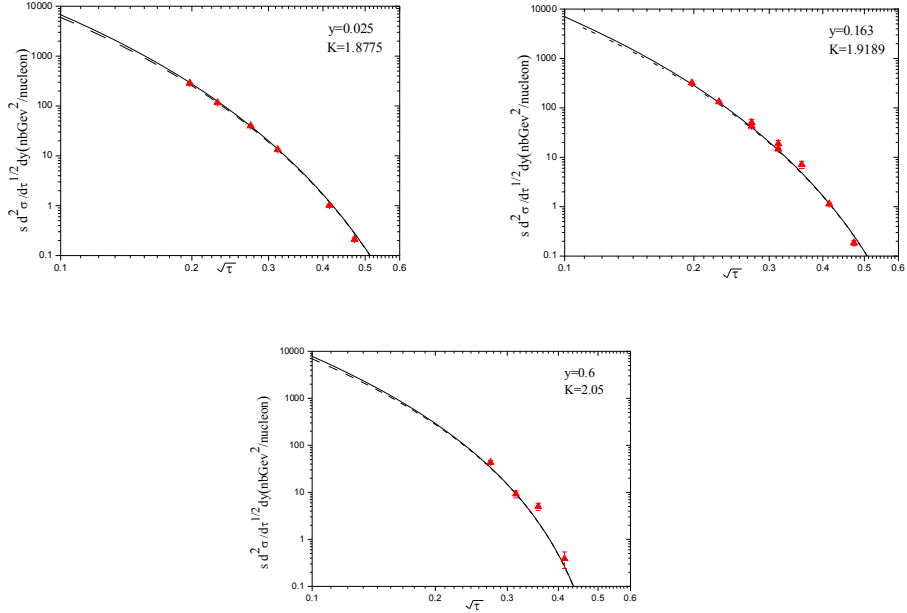

Figure 10: Dileptons production cross-section for p-Pt collision in terms of $\sqrt{\tau}$ according to Eq.13 for y = 0.025, 0.163, 0.6 and $\sqrt{s}$ = 38.8 GeV. The full curve is obtained by considering the pionic contribution and, the dash line is plotted without considering the pionic contribution. The used K factor has been indicated in each graph. The difference between full and dash lines is about 3 - 6% . The experimental results have been taken from [28].

**Conclusions**

In this study, by using the GRV's quarks distribution functions of nucleons and pions, lepton and anti-lepton production cross section in p-Cu and p-Pt collision was investigated. For this purpose, the pionic contribution in the structure function of Cu and Pt nuclei and their corresponding EMC ratio was first calculated. The results showed that for x=0.15 the pionic contribution increased the structure function of $^{63}$Cu and $^{195}$Pt to 7.9% and 8.8%, respectively. Also, for x=0.15, and by considering pionic contribution, the EMC ratio of $^{63}$Cu and $^{195}$Pt nuclei was increased to 5.44% and 6.28%, respectively. The results showed that pion cloud increased the dileptons production crosssection up to about 3- 8% in p- Cu collision and about 3- 6% in p- Pt collision. Comparison of the extracted results with the experimental data showed that these percentages for pionic contribution are appropriate.